\documentstyle[11pt,newpasp,twoside,epsf]{article}
\def\edcomment#1{\iffalse\marginpar{\raggedright\sl#1\/}\else\relax\fi}
\reversemarginpar

\begin{document}
\title{Mapping Metal--Enriched High Velocity Clouds to
Very Low H~{\sc i} Column Densities}
\author{Chris Churchill, Jane Charlton, Joe Masiero}
\affil{The Pennsylvania State University}


\begin{abstract}

Our galaxy is the nearest  known quasar absorption line system, and it
uniquely provides  us with an  opportunity to probe multiple  lines of
sight   through  the  same   galaxy.   This   is  essential   for  our
interpretations  of  the  complex   kinematic  profiles  seen  in  the
{\hbox{{\rm Mg}\kern 0.1em{\sc ii}}}  absorption due to lines of sight
through intermediate redshift galaxies.   The Milky Way halo has never
been  probed for  high  velocity clouds  below  the $21$-cm  detection
threshold       of        $N({\hbox{{\rm       H}\kern       0.1em{\sc
i}}})\sim10^{18}$~{\hbox{cm$^{-2}$}}.  Through a survey of {\hbox{{\rm
Mg}\kern 0.1em{\sc ii}}} absorption  looking toward the brightest AGNs
and  quasars, it  will  be possible  to  reach down  a  few orders  of
magnitude in  {\hbox{{\rm H}\kern 0.1em{\sc i}}}  column density.  The
analogs to  the high velocity  components of the  {\hbox{{\rm Mg}\kern
0.1em{\sc  ii}}}  absorption  profiles  due to  intermediate  redshift
galaxies should  be seen.  We  describe a program we  are undertaking,
and present some preliminary findings.

\end{abstract}

\section{Introduction}

Absorption  lines  in  quasars   provide  a  sensitive  probe  of  the
ionization, chemical, and kinematic conditions in galaxies.  To date, the
{{\rm Mg}\kern  0.1em{\sc ii}~$\lambda\lambda 2976,  2803$} doublet is
most  noted  as being  associated  with  galaxies  (e.g.\ Bergeron  \&
Boiss\'{e} 1991; Steidel, Dickinson, \&  Persson 1994) and is the most
well  studied at  resolutions  where the  kinematics  can be  resolved
within the  galaxy ``halos'' themselves (e.g.\  Churchill, Steidel, \&
Vogt 1996; Churchill et~al.\ 2000a; Churchill et~al.\ 2000b).

In a study of the kinematics, Churchill \& Vogt (2001) have shown that
most  all {\hbox{{\rm  Mg}\kern  0.1em{\sc ii}}}  systems exhibit  low
column  density, high  velocity ``subsystems''.   An example  of these
systems in shown in Figure~1.  As can be seen, there are a significant
number    of    subsystems    with    velocities    in    excess    of
100~{\hbox{km~s$^{-1}$}} from  the strongest, or  dominant, subsystem.
A detailed study of the complex  system at $z=0.93$ in the spectrum of
PG 1206+459  has revealed that these subsystems  have neutral hydrogen
column  densities  of  $N({\hbox{{\rm  H}\kern  0.1em{\sc  i}}})  \sim
10^{15.5}$~{\hbox{cm$^{-2}$}} (Churchill \& Charlton 1999).

\begin{figure}[th]
\plotfiddle{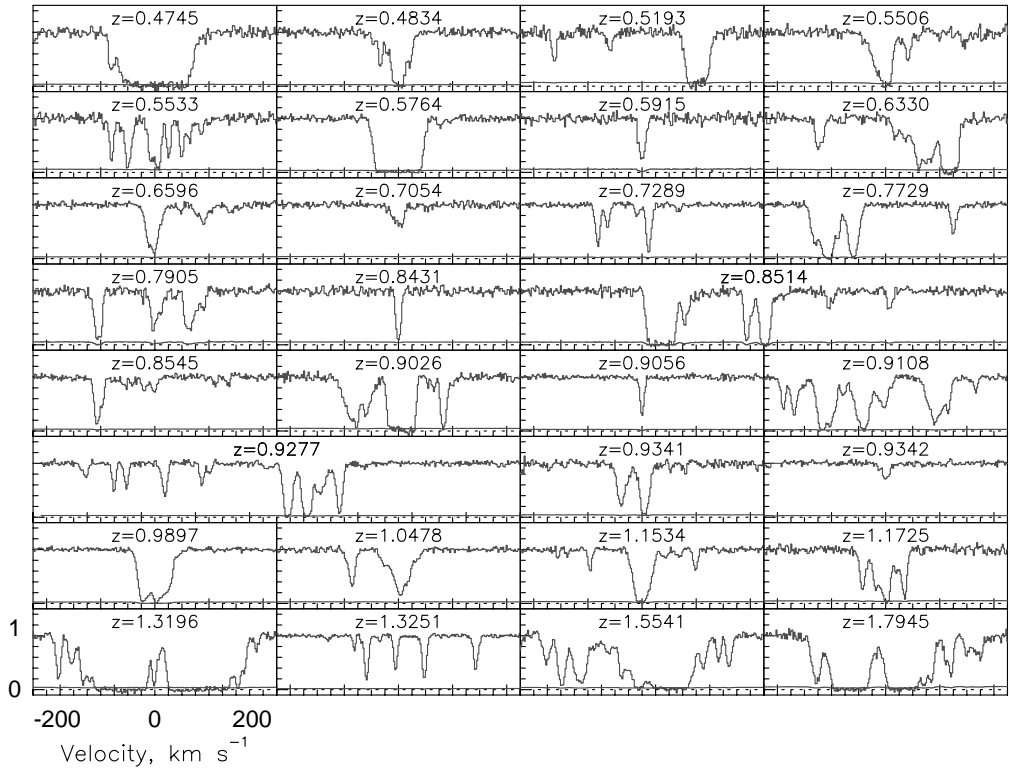}{3.75in}{0}{85.}{85.}{-275}{-330}
\caption{Keck/HIRES  spectra of  {\hbox{{\rm Mg}\kern  0.1em{\sc ii}}}
$\lambda 2796$  absorption profiles shown in  rest--frame velocity The
redshifts  of  absorption are  given  in  each  panel, which  shows  a
velocity range of 480~{\hbox{km~s$^{-1}$}}.}
\end{figure}

In  this contribution,  we argue  that {\hbox{{\rm  Mg}\kern 0.1em{\sc
ii}}} absorption line data through the extended regions of the Galaxy,
in as  many directions as feasible, are  requisite for ``calibrating''
the signatures  of the  various physical processes  that give  rise to
complex {\hbox{{\rm  Mg}\kern 0.1em{\sc ii}}}  gas kinematics observed
at  large   galactocentric  distances  around   intermediate  redshift
galaxies.   E230M/STIS,  high--resolution  spectra  of  {\hbox{{\rm
Mg}\kern 0.1em{\sc  ii}}} along  extragalactic lines of  sight through
the  Galactic halo  would provide  a sizable,  uniform,  database with
which  we  could  search  for  {\hbox{{\rm  Mg}\kern  0.1em{\sc  ii}}}
absorption  selected high--velocity  clouds  (HVC) in  the Galaxy  and
Local Group.  This would allow us  to map the sky distribution of HVCs
to {\hbox{{\rm  H}\kern 0.1em{\sc i}}} column  densities three decades
below present limits.

\section{HVCs: What are They?}

The  intermediate redshift  data, because  they allow  a  larger scale
perspective of  the spatial and  kinematic distribution of  gas around
galaxies provide statistical constraints on Galaxy data.  For example,
the nature, origin, and evolutionary role of the high--velocity clouds
(HVCs,  gas clouds  that depart  from galactic  rotation by  more than
100~{\hbox{km~s$^{-1}$}})  surrounding the  Galaxy remain  elusive and
are a  matter of current debate  (Blitz et~al.\ 1999;  Braun \& Burton
1999; Charlton, Churchill, \& Rigby  2000).  It is believed that their
role is  central to the cosmic  evolution of galaxies  and possibly to
galaxy  groups in  general,  but locally  it  still remains  uncertain
whether they  are strictly a  Galactic phenomenon or if  a substantial
number are deployed throughout  the Local Group.  Using the statistics
from  intermediate  redshift   {\hbox{{\rm  Mg}\kern  0.1em{\sc  ii}}}
absorption systems, Charlton et~al.\ showed that HVC--like material is
probably {\it not\/} distributed uniformly throughout galaxy groups.

From the study of quasar absorption lines, we know more globally about
the   {\it  extended\/}   spatial  and   kinematic   distributions  of
metal--enriched,  low ionization,  low  {\hbox{{\rm H}\kern  0.1em{\sc
i}}}  column density  ($10^{16} \leq  N({\hbox{{\rm  H}\kern 0.1em{\sc
i}}}) \leq  10^{18}$~{\hbox{cm$^{-2}$}}) gas in  intermediate redshift
($0.5 \leq z \leq 1.5$) galaxies  than we do about the Galaxy or Local
Group.  For  historical reasons,  apart from a  very few  works (e.g.\
Bowen, Blades,  \& Pettini 1995;  Savage, Sembach, \& Lu  1997; Savage
et~al.\ 2000), quasar absorption line  studies of the outer regions of
the Galaxy have  been limited mostly to low  resolution spectra, which
do not  reveal cloud--cloud velocity splittings  and column densities.
This  is because  the key  transitions for  study, such  as  the {{\rm
Mg}\kern  0.1em{\sc ii}~$\lambda\lambda 2976,  2803$} doublet,  are in
the Near UV (NUV).

At intermediate redshift the  transitions are observed in the optical,
where  large aperture  ground  based telescopes  with high  resolution
spectrographs are a powerful  resource and where $\sim 100$ high--flux
quasar are available for study.  Thus, a sizable, uniform database has
been acquired from which case--by--case studies and overall statistics
can  be  compiled (Churchill  \&  Vogt  2001).   

Most  absorbers are  characterized by  a dominant  kinematic subsystem
with $W_{r}(2796) > 0.2$~{\AA}  and velocity spreads ranging from $10$
to  $50$~{\hbox{km~s$^{-1}$}} in proportion  to the  system equivalent
width.  Weak, narrow, ``high  velocity clouds'' are observed in almost
every  intermediate  redshift  {\hbox{{\rm  Mg}\kern  0.1em{\sc  ii}}}
absorber,  and  these  are  inferred to  have  $N({\hbox{{\rm  H}\kern
0.1em{\sc  i}}})   \leq  10^{16.5}$~{\hbox{cm$^{-2}$}}  (Churchill  \&
Charlton  1999).  Weak  subsystems are  asymmetrically  distributed in
velocity such that  they are either all blueshifted  or all redshifted
with respect  to the dominant subsystem.  This  implies, that although
the ``kinematic morphologies'' vary greatly on a case--by--case basis,
a given line of sight is apparently probing a well defined spatial and
kinematic structure.

\begin{figure}[ht]
\plotfiddle{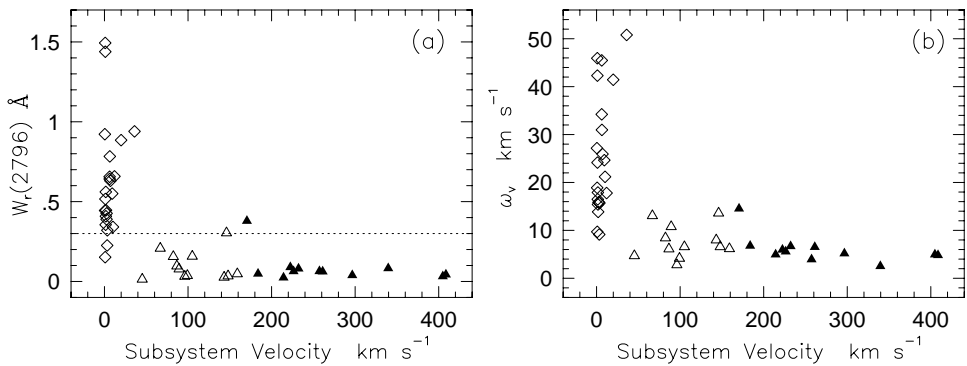}{1.9in}{0}{55.}{55.}{-220}{-150}
\caption{The rest--frame equivalent widths (a) and the velocity widths
(b) of  {\hbox{{\rm Mg}\kern 0.1em{\sc ii}}}  $\lambda 2796$ kinematic
subsystems.  Open  diamonds are  the dominant, strong  subsystem, open
triangles are those with $40 \leq v \leq 160$~{\hbox{km~s$^{-1}$}} and
solid triangles have $v > 160$~{\hbox{km~s$^{-1}$}}.}
\end{figure}

In Figure~2, we  show the rest--frame equivalent widths  (panel a) and
the  velocity  widths (panel  b)  of  the  kinematic subsystems.   The
velocity  zero points  are  given by  the  optical depth  mean of  the
profiles, which  lie within  the dominant subsystem  (open triangles).
Almost all of  the smaller subsystems have  equivalent widths less
than   0.3~{\AA}.    Note   that   the  velocities   extend   out   to
400~{\hbox{km~s$^{-1}$}}  and have very  narrow  velocity widths.
These weak, narrow clouds  have remarkable properties.  They are often
rich  in  {{\hbox{{\rm   Fe}\kern  0.1em{\sc  ii}}}}  absorption,  have
very--near     solar     metallicities,     densities     of     $\sim
0.1$~{\hbox{cm$^{-3}$}}, sizes of 10s of parsecs, and total gas masses
of 10s of solar masses (see Rigby , Charlton, \& Churchill 2001).

Based  upon line--of--sight number  densities, these  small subsystems
are inferred  to outnumber galaxies by  a factor of a  million to one
(Churchill   et~al.\  1999;  Rigby   et~al.\  2001;   Charlton,  these
proceedings).   Furthermore, the equivalent  width distribution  has a
cut  off below $W_{r}(2796)  = 0.08$~{\AA}  (Churchill \&  Vogt 2000),
implying a  paucity of  clouds with $N({\hbox{{\rm  Mg}\kern 0.1em{\sc
ii}}})$  smaller  than  $10^{12.3}$~{\hbox{cm$^{-2}$}}.  It  would  be
interesting  to see  if this  cut off  is present  in the  Galaxy.  An
examination  of cloud  destruction mechanisms  reveals that  that each
mechanism  (Rayleigh--Taylor   instability,  cloud--cloud  collisions,
Kelvin--Helmholtz instability,  and cloud evaporation  or condensation)
favors longer lifetimes for small  clouds in smaller mass halos and at
large  galactocentric  distances.   This  by  no  means  explains  the
differences in the equivalent  width distribution of moderate and high
velocity  subsystems   and  weak  systems   for  small  $W_{r}(2796)$.
However,  it is suggestive  that small  $W_{r}(2796)$ clouds  are less
favored in larger mass halos.   Of course, this assumes the ionization
conditions  and/or metallicities  do not  systematically  vary between
weak systems  and weak subsystems.  As stated  above, current evidence
is that they do not.

These results implies a  huge population around galaxies of low--mass,
low--ionization, small, high--velocity  clouds that have been enriched
by late stages of star formation.

\section{The Untapped Galactic Rosetta Stone and HVCs}

The sky  locations and kinematics  of Galactic HVCs remain  limited to
higher   {\hbox{{\rm   H}\kern    0.1em{\sc   i}}}   column   density,
$N({\hbox{{\rm        H}\kern        0.1em{\sc       i}}})        \geq
10^{18}$~{\hbox{cm$^{-2}$}},  based upon  21--cm  emission sensitivity
limitations.   This  limits knowledge  of  the $N({\hbox{{\rm  H}\kern
0.1em{\sc    i}}})$   mass    function   to    greater    than   $\sim
10^{7}$~M$_{\odot}$.   Wakker \& van  Woerden (1997)  extrapolated the
21--cm maps  and concluded that the  sky covering factor  of HVCs with
$N({\hbox{{\rm  H}\kern 0.1em{\sc i}}})  = 10^{17}$~{\hbox{cm$^{-2}$}}
is 30--60\%.   It has required  Herculean feats to probe  further down
the mass function, with only limited success; it still remains unknown
whether  the mass  function  rises  with decreasing  mass  or is  flat
(however,  see other  contributions to  these proceedings).   This has
profound implications  for understanding  if known HVCs  represent the
tip  of  the  high--mass   ``ice--berg''  of  a  heretofore  invisible
reservoir of  metal--enriched gas  clouds and for  understanding their
origin and role in Galactic and/or Local Group evolution.

\begin{figure}[pth]
\plotone{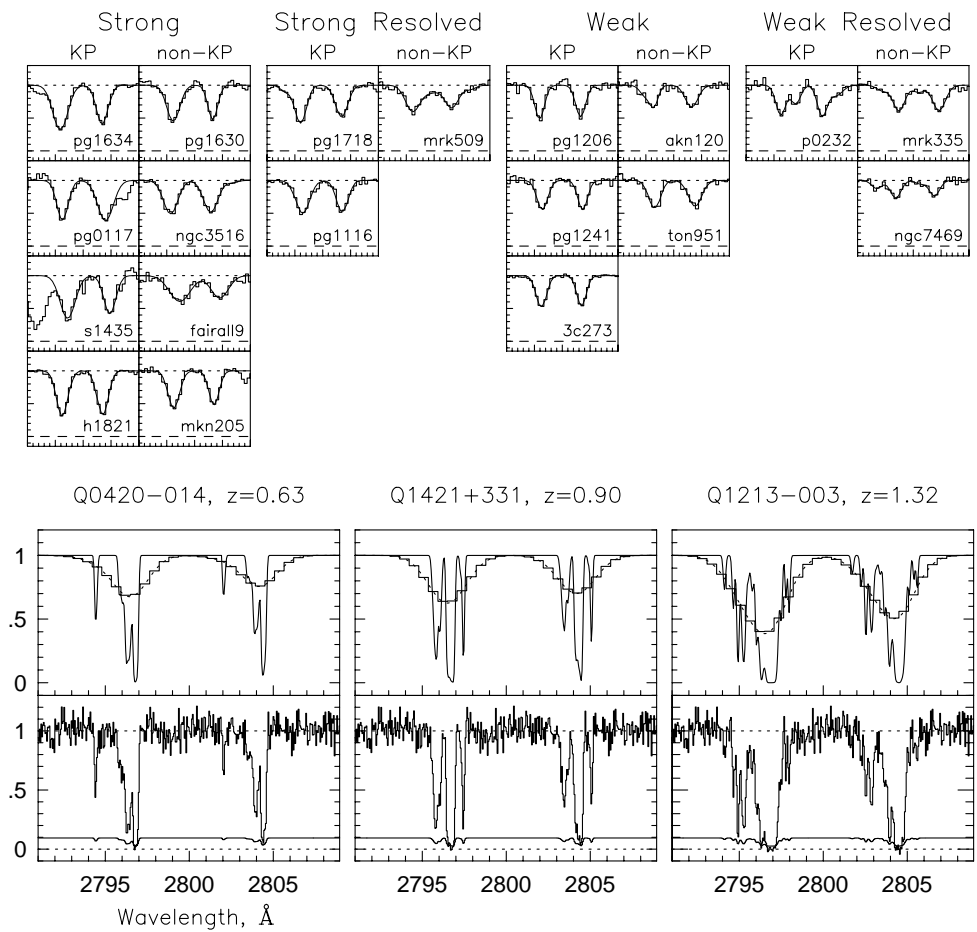}
\caption{The top  panels are  FOS/HST spectra of  Galactic 
{{\rm Mg}\kern 0.1em{\sc ii}~$\lambda\lambda 2976, 2803$}
doublets  taken from  Savage  et~al.\  (2000); the  lines  of sight  are
organized by their classifications.   ``Resolved'' profiles are due to
HVCs along the  line of sight.  The lower panels  show examples of how
observed {\hbox{{\rm Mg}\kern 0.1em{\sc  ii}}} systems with low column
density HVCs  would look in  the Savage et~al.\ spectra.   The departure
from unresolved lines increases to the right.}
\end{figure}

For the Galaxy, {\it HST\/} is  required for the NUV.  The problem has
been that the  photon collecting power of {\it  HST\/} has limited the
number  of  quasars for  which  high--resolution  spectroscopy can  be
acquired to the few very brightest  in order to avoid large amounts of
telescope time.  This means  we currently have no uniform, statistical
database nor  knowledge of  whether counterparts to  the intermediate,
low  $N({\hbox{{\rm H}\kern  0.1em{\sc  i}}})$, narrow  high--velocity
clouds are present  in the outer regions of the  Galaxy.  The irony of
all this is  that detailed knowledge of low  column density gas around
the Galaxy is the key  to interpreting the intermediate redshift data.
The Galaxy  is literally the ``Rosetta Stone''  of metal--line systems
because  proximity allows  the  absorption line  data  to be  directly
compared  with  the  structures   associated  with  the  gas  and  the
orientation of the line of sight through the Galaxy disk/halo.

\section{HVC Studies... Yes. But, How to Compare to Mg II?}

A  great deal  of effort  has been  directed toward  understanding the
kinematics,   ionization  and  chemical   conditions  of   HVCs  using
absorption  lines  in  the spectra  of  halo  stars  and of  the  very
brightest  quasar/AGN observed  from both  the ground  and  from space
(e.g.\  Savage  \& Sembach  1996,  and  references  therein), and  the
results  have  included the  dust  properties, abundances,  ionization
conditions, kinematics (the literature is far to vast to review here).
A  recent interesting  result,  for  example, is  that  some HVCs  are
composed  of ``cloudlets''  with  densities and  temperatures not  too
dissimilar to  those seen in  {\hbox{{\rm Mg}\kern 0.1em{\sc  ii}}} at
intermediate redshifts (Lehner et~al.\ 1999).

The problem is that virtually none  of the Galactic HVC studies can be
directly compared and contrasted  to the sizable intermediate redshift
sample  of absorption  line data.   From the  ground,  the {\hbox{{\rm
Ca}\kern  0.1em{\sc  ii}}}, {\hbox{{\rm  Na}\kern  0.1em{\sc i}}},  and
{\hbox{{\rm Ti}\kern 0.1em{\sc ii}}} transitions have been observed at
high resolution.   Though these studies have the  sensitivity to probe
low column  {\hbox{{\rm H}\kern 0.1em{\sc  i}}}, they sample  the most
neutral,  higher density  components  of  the gas,  which  are not  as
ionized as {\hbox{{\rm  Mg}\kern 0.1em{\sc ii}}}--selected gas clouds.
Almost exclusively, absorption line studies of HVCs using {\it HST\/}
have focused on the high ionization transitions of {\hbox{{\rm C}\kern
0.1em{\sc  iv}}},  {\hbox{{\rm
N}\kern 0.1em{\sc v}}}, and {\hbox{{\rm O}\kern 0.1em{\sc vi}}}.

Savage et~al.\ (2000) published an {\it HST\/} Key Project (KP) search 
for  {\hbox{{\rm  Mg}\kern 0.1em{\sc  ii}}}  HVC  absorption along  71
extragalactic      sight      lines      using     low      resolution
($v=230$~{\hbox{km~s$^{-1}$}})  FOS/{\it HST\/}  spectra.   Though the
resolution is too poor to  directly resolve HVC absorption, they found
41  sight   lines  with  {\hbox{{\rm  Mg}\kern   0.1em{\sc  ii}}}  HVC
absorption based  upon the equivalent  widths and profile  shapes.  In
Figure~3, we  show the  KP spectra and  the Savage  class, ``strong'',
``strong resolved'',  ``weak'' and ``weak resolved''.   We include the
archived  FOS profiles  and  have determined  the  Savage classes  for
quasars that should be  observed with STIS.  The ``resolved'' profiles
are expected  to have strong {\hbox{{\rm Mg}\kern  0.1em{\sc ii}}} HVC
absorption.  The  bottom panels of  Figure~3 show a  direct comparison
between  three intermediate  redshift  {\hbox{{\rm Mg}\kern  0.1em{\sc
ii}}} systems (noiseless models) and  how they would look as noiseless
FOS spectra.

Bowen  et~al.\  (1995)  have  investigated  high--resolution  Galactic
{\hbox{{\rm Mg}\kern 0.1em{\sc ii}}}  absorption along seven (7) lines
of  sight using  GHRS.  They  found a  {\hbox{{\rm  Mg}\kern 0.1em{\sc
ii}}} HVC in only one sight line, but their detection threshold ranged
from   $0.5  \leq  W_{r}(2796)   \leq  0.7$~{\AA}   ($2~\sigma$).   By
comparison, the  intermediate redshift {\hbox{{\rm  Mg}\kern 0.1em{\sc
ii}}} data  are sensitive to $W_{r}(2796)  = 0.02$~{\AA} ($5~\sigma$).
Given the power--law  slope of $\simeq -1.5$ for  the equivalent width
distribution   of  narrow,   {\hbox{{\rm   Mg}\kern  0.1em{\sc   ii}}}
``high--velocity clouds'' at intermediate redshifts (Churchill \& Vogt
2001), it would  not be expected for there to  be a substantial number
of relatively strong {\hbox{{\rm  Mg}\kern 0.1em{\sc ii}}} HVCs in the
small Bowen et~al.\ sample.

\section{The Galactic Sky at High Resolution}

In Figure~4,  we show an Aitoff  projection of the  location of viable
halo sight lines.   They are viable in that the  fluxes of the quasars
or AGN are high enough that reasonable signal to noise can be obtained
in a reasonable integration time with HST.  The locations of the major
HVCs  are  shown schematically.   Note  that  the  viable sight  lines
represent a fairly isotropic coverage of the halo, sampling both large
HVC complexes and regions far from HVCs.

\begin{figure}[t]
\plotfiddle{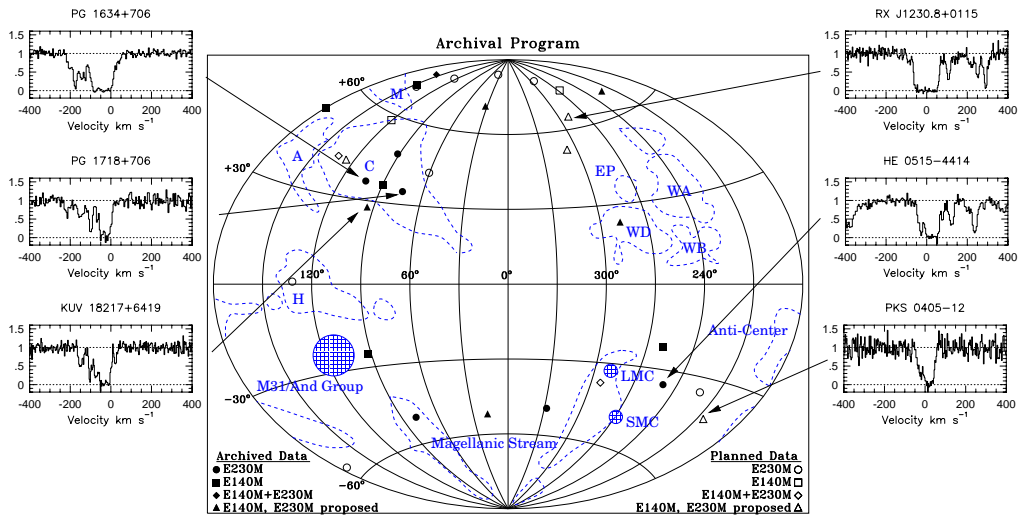}{3.5in}{-90}{70.}{70.}{-198}{200}
\caption{Six   absorption  profiles   showing   HVC  material,   three
associated with Complex C and  three far from known HVCs.  The legends
in the Aitoff diagram give  the available spectra from the HST archive
and also  the desired  (planned) lines of  sight. (The  unpublished PG
1634+706 spectrum is courtesy of Buell Jannuzi.)}
\end{figure}

Three spectra  probing Complex C are  shown to the left  of the Aitoff
diagram.  All  three lines  of  sight  are  consistent with  the  bulk
velocity  of  complex  C (about  $-120$~{\hbox{km~s$^{-1}$}}).   Note,
however, the  very different  fine structure in  the velocity  and how
this is different along each line of sight.

As a further example of the absorption line spectra, three systems are
shown    in   Figure~5.     Note    the   weak    narrow   cloud    at
$v=+120$~{\hbox{km~s$^{-1}$}}  toward  NGC   4151,  which  probes  the
Galactic Pole away  from the large HVC complexes!  In  this case it is
difficult to know if this Galactic in origin or intrinsic to NGC 4151,
which has a  very low redshift.  This raises  the further concern that
the  selected  sight  lines  should have  significant  redshifts.   In
Figure~5,  we also  show the  {{\hbox{{\rm Fe}\kern  0.1em{\sc ii}}}},
{\hbox{{\rm  Mn}\kern   0.1em{\sc  ii}}},  and   {\hbox{{\rm  Mg}\kern
0.1em{\sc  i}}} transitions.   {{\hbox{{\rm Fe}\kern  0.1em{\sc ii}}}}
allows  confirmation  of  weak  {\hbox{{\rm Mg}\kern  0.1em{\sc  ii}}}
clouds and {\hbox{{\rm Mn}\kern 0.1em{\sc ii}}} allows structure to be
discerned in  the saturated regions of  {\hbox{{\rm Mg}\kern 0.1em{\sc
ii}}} and {{\hbox{{\rm Fe}\kern 0.1em{\sc  ii}}}} so that we can Voigt
profile the spectra precisely as we analyzed the intermediate redshift
data (Churchill \& Vogt 2001).

\begin{figure}[ph]
\plotfiddle{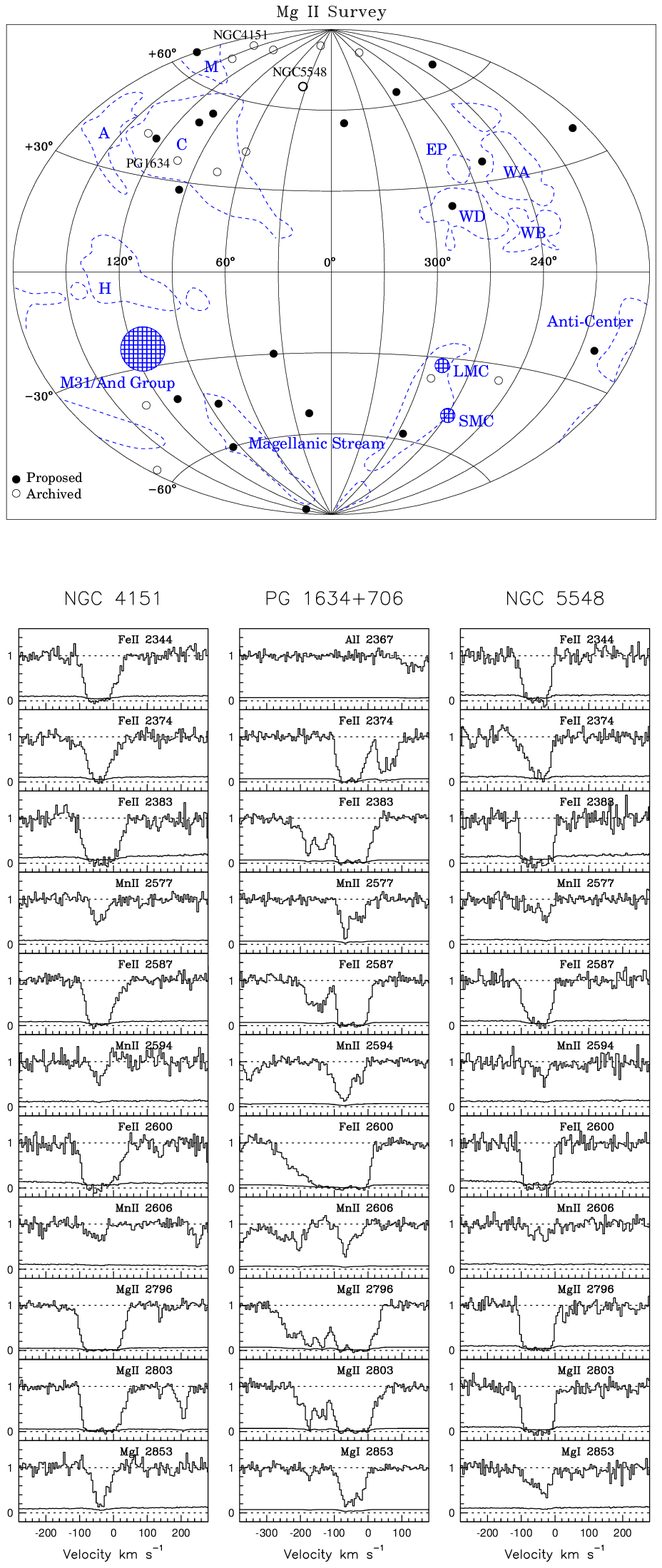}{7.5in}{0}{90.}{90.}{-128}{-10}
\caption{An example of the many transitions available for study of
ionization conditions, abundances and kinematics.  The three lines of
sight are labeled in the Aitoff diagram. (The unpublished PG 1634+706
spectrum is courtesy of Buell Jannuzi.)}
\end{figure}

\pagebreak

\end{document}